\documentclass[aps,groupedaddress,a4paper,floatfix,twocolumn,footinbib]{revtex4}
\usepackage{pslatex}
\usepackage{amsfonts}
\usepackage{amssymb}
\usepackage{mathrsfs}
\usepackage{ifthen}
\usepackage[mathscr]{euscript}
\usepackage{graphicx}
\usepackage{fancyhdr}
\usepackage[hypertex=true]{hyperref} 

\begin{document}

\title{
An optimised algorithm for ionized impurity scattering in 
Monte Carlo simulations
}

\author{W.Th. Wenckebach}
\email{wenckebach@tn.tudelft.nl}
\affiliation{
Department of Applied Physics, Technical University Delft, 
Lorentzweg 1, 2628 CJ DELFT, The Netherlands. 
}
\author{P. Kinsler}
\email{Dr.Paul.Kinsler@physics.org}
\altaffiliation[New address:]{
Department of Physics, Imperial College,
  Prince Consort Road,
  London SW7 2BW, 
  United Kingdom.
}
\affiliation{
Department of Applied Physics, Technical University Delft, 
Lorentzweg 1, 2628 CJ DELFT, The Netherlands. 
}

\lhead{
\includegraphics[height=5mm,angle=0]{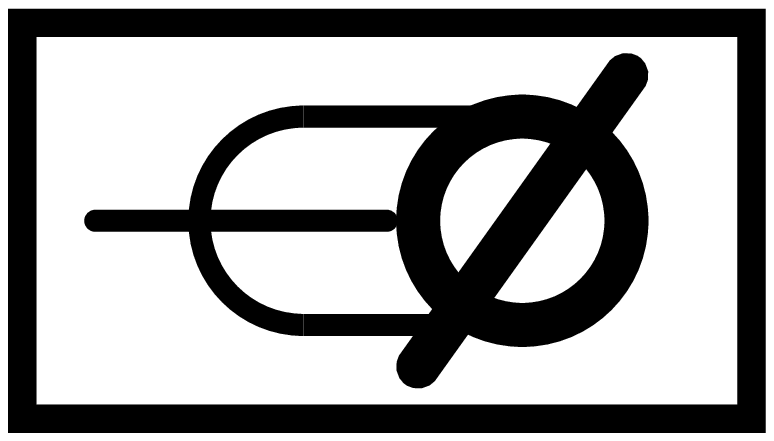}~~
OPTIIM}
\chead{
\href{mailto:wenckebach@tn.tudelft.nl}{wenckebach@tn.tudelft.nl}\\
\href{http://terahertz.tn.tudelft.nl/}{http://terahertz.tn.tudelft.nl/}
}
\rhead{
\href{mailto:Dr.Paul.Kinsler@physics.org}{Dr.Paul.Kinsler@physics.org}\\
\href{http://www.kinsler.org/physics/}{http://www.kinsler.org/physics/}
}

\begin{abstract}

We present a new optimised model of Brookes-Herring ionized impurity
scattering for use in Monte Carlo simulations of semiconductors. 
When implemented, it greatly decreases the execution time needed for
simulations (typically by a factor of the order of 100), and also
properly incorporates the great proportion of small angle scatterings
that are neglected in the standard algorithm.  It achieves this
performance by using an anisotropic choice of scattering angle which
accurately mimics the true angular distribution of ionized impurity
scattering.

\end{abstract}

\maketitle
\thispagestyle{fancy}



%



{\em
This paper was published as 
Computer Physics Communications {\bf 143}, 136 (2002).
}

\section{Introduction}

Ionized impurity scattering\cite{Ridley-QPS} dominates the transport
properties of highly doped semiconductors, but accurate modeling of
this scattering process is difficult because of the long range of the
Coulomb force. The method in general use is that by Brookes-Herring,
which is regarded as sufficiently accurate when simulating carrier
mobilities in semiconductors.  However, Brookes-Herring scattering is
very anisotropic, so the standard overestimation-rejection
algorithm\cite{Jacoboni-R-1983} for implementing this in Monte Carlo
simulations is hugely inefficient.  

The overestimation-rejection algorithm works as follows.  The
overestimated scattering rate is used to determine the free flight
time of the particle, and at the end of every period of free flight a
check is made to see if a scattering event has occurred.  This, of
course, leads to more checks being made than needed, but this problem
is unavoidable since we cannot know in advance what the exact rate,
for whatever final state that eventuates, will be.  Each check
involves three stages: selection of a possible final state, a
calculation of the scattering rate {\em into} that final state would
be, and a decision as to whether the scattering occurs or not.

This means that despite the fact that a scattering event does not
occur at the end of every free flight, we still have to calculate the
exact differential scattering rate for some particular choice of
scattering event.  If the overestimation is too great, this means the
program will calculate far more differential scattering rates than
strictly necessary; and then if this calculation is computationally
expensive, the simulation will need to run for a much longer time. 
This is generally the case with ionized impurity scattering models in
semiconductors, whose isotropic selection of possible final states
means that overestimated rate must also be isotropic.  Since the
differential scattering rate is strongly peaked for small angles,
this means that the overestimated maximum scattering rate is far in
excess of the typical differential rate.

Our new algorithm has the angular dependence of ionized impurity
scattering built in, thus our overestimation becomes extremely
efficient (see e.g.  \cite{Konijn-RW-1993}).  The resulting code
therefore will run the same simulation faster than one with the
standard isotropic algorithm; or, alternatively, could run a much
more accurate simulation in a comparable time.  An early version of
this algorithm was used recently to obtain manageable execution times
in simulations of hot-hole lasers in III-V
semiconductors\cite{Kinsler-W-2001jap}.

The paper is organised as follows: section \ref{Siim} describes the
different algorithms; section \ref{Sresults} describes their
comparative performance, and finally, in section \ref{Sconclusions}
we present our conclusions.


\section{Ionized Impurity Scattering}\label{Siim}

Here we are interested in optimising the algorithm for the
Brookes-Herring model of ionized impurity scattering.  This looks at
the likelihood of the particle of interest (electron or hole) being
deflected by the screened Coulomb field from a charged
impurity\cite{Ridley-QPS}.  The Brookes-Herring rate for ionized
impurity scattering is

\begin{equation}
W_{diff} \; = \; \frac{2 \pi}{\hbar} \; 
          \frac{ n_{ii} e^4}{\varepsilon_0^2 \varepsilon_r^2} \;
          \frac{1}{(q^2 + q_0^2)^2} \; N_d (E^m_f, \theta_f, \phi_f )
 \; 
          \left| \left< nk_i \mid mk_f \right> \right|^2
\label{e-Wdiff}
\end{equation}

Here $n_{ii}$ are the ionized impurity and hole (or electron) concentrations,
$q_0^2$ is the square of the minimum change of the $k$-vector, $k_B$
is Boltzman's constant, $T$ the temperature, $\varepsilon_0
\varepsilon_r$ the permitivity, $e$ the electron charge, $i$ is the
number of the initial band, $f$ is the number of the final band.  The
so-called overlap factor is the square of the scalar product of the
eigenvectors corresponding to the initial and final state
respectively, and is $\left| \left< nk_i \mid mk_f \right> \right|^2$.

Note in particular that the rate $W_{diff}$ is proportional to

\begin{equation}
     \frac{1}{\left (q^2 + q_0^2 \right)^2} \; = \;
     \frac{1}{\left( \left|{\bf k}_f - {\bf k}_i \right|^2 
                     + q_0^2 \right)^2 } 
\label{e-Wdiffprop}
\end{equation}

We would usually assume that $q_0$ is just the inverse of 
the Debeye screening length, which is given by

\begin{equation}
      q_{d}^2 \; = \; \frac{e^2 N_{ii}}{
             \varepsilon_0 \varepsilon_r \; k_B T}  
,
\label{e-Debeye}
\end{equation}

although in subsection \ref{ss-ModAniAlg} we introduce an additional
contribution $k_{min}$, where $q_{0}^2=q_{d}^2+k_{min}^2$.

The difficulty with ionized impurity scattering is that the
scattering rate becomes very large for small values of $q = |{\bf
k}_f - {\bf k}_i|$.  We can distinguish two situations leading
scattering rates that are very large compared to the average
scattering rate:

\begin{description}

\item[ (1) ] small angle scattering, i.e. $q$ is small compared to
the absolute value of the incoming wave vector $k_i = \left| {\bf
k}_i \right|$.  Note that $k_i$ is of the same order of magnitude as
the absolute value of the outgoing wave vector $k_f = \left| {\bf
k}_f \right|$.

\item[ (2) ] low hole or electron energy, i.e. where $k_i$ is small 
by itself.

\end{description}

\subsection{Standard Isotropic Algorithm}
\label{ss-StdIsoAlg}

The great variation in the scattering rate by angle and wave vector
means that it is not straightforward to efficiently model the
scattering process.  If we take the simple approach of an isotropic
choice of scattering angle, the usual algorithm  overestimates the
differential scattering rate $W^{diff}_p (E_k(t),\theta ,\phi)$ by
its maximum value $W^{diff}_{max}$.  Therefore our Monte Carlo
program will select an ionized impurity scattering for any particle
at the overestimated rate $W^{diff}_{max}$, then reject those for
which the combination of exact scattering rate $W^{diff}_p
(E_k(t),\theta ,\phi)$ and choice of random number do not lead to a
scattering.  

To choose a possible final state for an ionized impurity scattering
process, we start by choosing a random direction for ${\bf k}_f$ with
all possible final directions have the same statistical weight
$f(\theta, \phi) = 1/(4\pi)$.  Once we have chosen this direction, we
can then work out the final $k$ vector, and hence the true scattering
rate.  We compare this true rate to the overestimated differential
scattering rate, and use the normal rejection procedure to determine
statistically whether the scattering has occurred or not.

For ionized impurity scattering $W^{diff}_{max}$ is much and much
larger than $W^{diff}_p (E_k(t),\theta ,\phi)$ for about any angle;
and also this  maximum value $W^{diff}_{max}$ depends sensitively on
the screening length.  This means that $W^{diff}_{max}$ is an
inefficient overestimation, and as a result we do many calculations
which do not lead to a scattering process.  This significantly
lengthens the execution times for our simulations.

\subsection{New Anisotropic Algorithm}
\label{ss-NewAniAlg}

We have improved significantly on the execution times of simulations
by introducing an anisotropic choice of the scattering angle, whilst
compensating for this by adjusting the final scattering rate.  The
angular dependence of the scattering rate is centered on the
direction of the initial $k$ vector, so it is useful to transform to
angular coordinates $(\theta, \phi)$ centered on this direction. For
this purpose we first define

\begin{equation}
      Q_1 \equiv \left| \frac{{\bf k}_f}{k_f} \; - \;
                        \frac{{\bf k}_i}{k_i} \right|
\end{equation}

so

\begin{equation} 
      Q_1 = 2 \sin (\theta / 2) 
          = \sqrt{2(1 - \cos \theta )}                
\end{equation}

where $\theta$ is the scattering angle.  We now introduce an
anisotropic weighing function $f(\theta, \phi)$ for the probability
of finding the direction of the final k-vector.  This same factor
will then be used to compensate for this weighting by subsequently
dividing the differential scattering by $f(\theta, \phi)$.  At first
sight one would be inclined to choose

\begin{equation}
      f(\theta, \phi) \; \propto \;
      \frac{1}{Q_1^4} \; = \;
      \frac{1}{16 \sin^4 (\theta / 2)} \; = \;
      \frac{1}{4 (1 - \cos \theta )^2}                         
\end{equation}

but this function cannot be normalised properly because the integral
of $f(\theta, \phi)$ over $\theta$ and $\phi$ diverges.  We could try
to solve this problem by assuming a minimum scattering angle
$\theta_0$, so we get a minimum value for $Q_1$ equal to $Q_0 = 2\sin
(\theta_0 / 2)$ and to choose for $Q_0$ e.g. the approximate
Debeye-value of Ridley\cite{Ridley-QPS}, we have

\begin{equation}
      (Q_0 k_i)^2 \; = \; q_0^2 
\end{equation}

We would then assume $\theta_0$ to be so small that we may assume the
absolute value of the incoming and outgoing wave vectors to be the
same.  However, excluding scatterings with $q < q_0$ leads to
erroneous results because it is exactly these small angle scatterings
which dominate ionized impurity scattering.  Test simulations confirm
that the normalisation procedure has an equally important influence
on the final result as the way screening is implemented.  

To avoid erroneous results we choose $f(\theta, \phi)$ to have the
Brookes-Herring shape (see eqn. \ref{e-Wdiffprop}) -- 

\begin{equation}
      f(\theta, \phi) \; \propto \;
      \frac{1}{\left( Q_1^2 + Q_0^2 \right)^2} \; = \;
      \frac{1}{\left( 4 \sin^2 (\theta / 2) + Q_0^2 \right)^2} \; = \;
      \frac{1}{\left( 2(1 - \cos \theta ) + Q_0^2 \right)^2}  
\end{equation}

The function can be normalised because 

\begin{eqnarray}
 \int_0^{2\pi} d\phi
 \int_0^{\pi} d\theta \;
 \sin \theta \;
 \frac{1}{\left( Q_1^2 + Q_0^2 \right)^2}
 & = &
 \frac{4\pi}{Q_0^2 \left( 4 + Q_0^2 \right)}
\end{eqnarray}

is finite.  As a result the normalised statistical weight of finding
a given direction is given by

\begin{equation}
f(\theta, \phi)
 \; = \; 
\frac{Q_0^2 \left( 4 + Q_0^2 \right)}{4\pi} \;
\frac{1}{\left( Q_1^2 + Q_0^2 \right)^2}
 \; = \; 
\frac{Q_0^2 \left( 4 + Q_0^2 \right)}
     {4\pi \left[ 2 \left( 1-\cos\theta \right) + Q_0^2 \right]^2}
\end{equation}

The next problem is to enter this statistical weight in the
algorithm.  To determine $\theta$, $\phi$ we choose two random
numbers, $r_1$ and $r_2$, between 0 and 1.  In the case of an
isotropic choice of direction we would have $f(\theta, \phi) = 1 /
4\pi$, and we would choose

\begin{eqnarray}
      r_1 & = & \frac{\phi}{2\pi}  \label{defr1}\\
      \rm{and} \; \;
      r_2 & = &  2\pi \left(1-\cos\theta\right) f(\theta, \phi)
.
\end{eqnarray} 


For our new algorithm, we choose $r_1$ as in eqn. (\ref{defr1}), and 
          
\begin{eqnarray}
  r_2 
& = & 
  \pi \left[ 2 \left( 1 - \cos\theta \right) + Q_0^2 \right] f(\theta, \phi)
  - \frac{Q_0^2}{4}
\end{eqnarray} 

The extra terms with $Q_0^2$ are needed to ensure that $r_2$ lies
between $0$ and $1$, so that $-1 < \cos \theta < 1$.  The weighed
random choice of $\phi$ and $\theta$ is now obtained by solving $\phi
= 2\pi \; r_1  $ and 

\begin{equation}
      \cos \theta 
  \; = \;
      1 \; + \; \frac{Q_0^2}{2} \left( 1 \; - \;
      \frac{4 + Q_0^2 }{4 r_2 + Q_0^2} \right)
\end{equation}

Note that $\cos \theta \rightarrow 1$ for $r_2 \rightarrow 1$, so
small angle scattering is not excluded.  

In the new algorithm we need to compensate for the higher occurrence
of small scattering angles caused by the introduction of a weighing
function $f(\theta, \phi ) \neq 1/4\pi$ by multiplying the
differential scattering rate by

\begin{equation}
      \frac{4\pi}{f(\theta, \phi )} \; = \;
      \frac{\left( Q_1^2 + Q_0^2 \right)^2 }{
      Q_0^2 \left( 4 + Q_0^2 \right) }   
\label{mdsr-by}
\end{equation}

We construct an overestimated differential scattering rate by
starting to look for the largest differential ionized impurity
scattering rate including the factor $f(\theta,\phi)$.  The overlap
factor is always less than one.  The term

\begin{equation}
    \frac{1}{(q^2 + q_0^2)^2}          
\end{equation}

can be overestimated by

\begin{equation}
    \frac{C}{(Q_1^2 + Q_0^2)^2 k_i^4}          
\end{equation}


where $C$ is equal to 1 if the band is isotropic, but may need to be
larger for anisotropic bands.  In practice,  $C$ needs to be
determined by computer experiment, which in our simulations was set
equal to 1.2.  There is also our new weighting factor
$f(\theta,\phi)$, which with the previous factor (eqn. 
\ref{mdsr-by}) may both be overestimated by

\begin{equation}
      \frac{\left( Q_1^2 + Q_0^2 \right)^2 }{
      Q_0^2 \left( 4 + Q_0^2 \right) } \; 
      \frac{C}{(Q_1^2 + Q_0^2)^2 k_i^4} 
  \; = \;    
      \frac{C}{Q_0^2 \left( 4 + Q_0^2 \right) } \; 
      \frac{1}{k_i^4} 
  \; \approx \;
      \frac{C}{q_0^2 \left( 4k_i^2 + q_0^2 \right) }
\end{equation}

Note that this is the point where choosing a good way to normalise
the weighing function $f(\theta, \phi )$ pays off.  The resulting
function becomes extremely simple!  Finally, the differential density
of states is replaced by the total density of states divided by
$4\pi$.  Thus the overestimated scattering rate is given by

\begin{equation}
W_{tot} \; = \; \frac{2 \pi}{\hbar} \; 
          \frac{ n_{ii} e^4}{\varepsilon_0^2 \varepsilon_r^2} \;
          \frac{1}{q_0^2 \left( 4k_i^2 + q_0^2 \right) }
          N_{tot} (E^n_i ) \; \frac{C}{4\pi}
\end{equation}

Almost always $k_i^4 \gg q_0^4$, so this overestimation is orders of
magnitude smaller than that from the standard isotropic algorithm,
leading to far fewer calculations of the differential scattering rate
that are followed by rejection.

\subsection{\bf Modified Anisotropic Algorithm}
\label{ss-ModAniAlg}

Ionized impurity scattering is an elastic process, which means that
small changes of the $k$-vector imply a small change of the state of
the hole or electron.  A minimum change $q_0$ of the $k$-vector
follows immediately from the implementation of the Debeye screening
through eqns. (\ref{e-Wdiff}) and (\ref{e-Debeye}).  The resulting
minimum value of $q_0$ is proportional to $\sqrt{N_{ii}}$, so it
becomes very small for low ionized impurity concentrations.  As a
result, despite the anisotropic scattering angle selection, the
fraction of overestimations and hence the execution times of the
simulation will still become very large for low concentrations.  We
can therefore improve the overestimation by introducing a second
criterion determining whether the change of the $k$-vector is small
or not.  Now the average energy of an electron or hole is given by

\begin{equation}
\frac{3}{2} \; k_BT \; = \; \frac{\hbar^2 <k^2>}{2m_e m^*}
\end{equation}

so the average value of $k^2$ is given by

\begin{equation}
<k^2> \; = \; \frac{3m_e m^* k_BT}{\hbar^2}  
\end{equation}

This induces us to propose that we consider $k$ small if its 
square is much smaller than this value.  So we set

\begin{equation}
k_{min}^2 \; = \; D \; \frac{3m_e m^* k_BT}{\hbar^2}
\end{equation}

where we choose $D$ to be a small number.  In practice we find that
$D = 0.01$ is acceptable.  We can now set the minimum change in wave
vector to be

\begin{equation}
q_0^2 \; = \; q_d^2 + k_{min}^2
\end{equation}

   Note that excluding scatterings with a `small' change $\Delta k$ of $k$
means skipping an infinite number of small angle scatterings.  However these
excluded scatterings have a negligible influence on the mobility since the
motion of the carrier is largely unaffected. E.g. if $\Delta k < c k_0$, where
$k_0^2/2m^* = k_b T$ and $c = 0.1$, then on the average such a scattering
changes the component of $k$ in    the direction of the electric field by less
than 1\%.  To verify this argument we repeated a number of the simulations as
varying $c$ from 0.1 to 1.6.  For $c \leq 0.3$ we saw no significant effect on
the calculated mobilities.


\section{Results}\label{Sresults}

To check the relative efficiency of these algorithms, we took an
existing Monte Carlo code (used in Dijkstra and
Wenckebach\cite{Dijkstra-W-1997jap}) and rewrote one of the
subroutines in two ways: first, in accordance with the isotropic
algorithm, and second, according to the modified anisotropic
algorithm.  We could then compare the performance of the two
algorithms by comparing the performance of the two codes compiled
with either of the two coded algorithms.  In all other aspects the
two codes were the same.  However, with the isotropic choice of the
scattering angle, small angle scatterings such that $\sin \theta <
q_0$ had to be excluded in order to let the program run within finite
time.  

Both the material parameters we used, and the general method followed
by the code were the same as those of Hinckely and Singh
\cite{Hinckley-S-1990,Hinckley-S-1994}.  The codes were compiled with
GNU g77 and run on a 500 MHz Pentium III running RedHat Linux 6.2.
The calculation was of the drift velocities of holes in unstrained
silicon, where the magnetic field was equal to zero, the electric
field equal to 5 kV/cm and the temperature 300 K.  Only the
$k$-vector was integrated during free flight.  Each simulation
consisted of either 25 blocks of 1000 real scatterings each or of 2
blocks of 100 real scatterings.  Only in the former set of
simulations could mobilities be determined.

\begin{table}[h]
\caption{
A comparison of execution times and mobilities for the isotropic and 
anisotropic algorithms.  The material system was unstrained silicon
with and electric field equal to 5 kV/cm and a temperature of 300 K.
Note that in almost all cases, the simulations averaged over $25 
\times 1000$ scatterings  take hours using the isotropic algorithm, 
but only minutes using the anisotropic one.
}
\begin{center}
\begin{tabular}{|l||c|r|c||c|r|}
\hline
\multicolumn{1}{|c|}{Code:}     &
            \multicolumn{3}{|c|}{isotropic choice of $\theta$} &
            \multicolumn{2}{|c|}{anisotropic choice of $\theta$} \\
\hline
Scatters.  & \multicolumn{2}{|c|}{25 $\times$ 1000} & 2 $\times$ 100 &
            \multicolumn{2}{|c|}{25 $\times$ 1000} \\
\hline
Conc.     & Exec. time & Mobility  & Exec. time & 
            Exec. time & Mobility  \\
cm$^{-3}$ & hh.mm.ss   & cm$^2$/Vs & hh.mm.ss   & 
            hh.mm.ss   & cm$^2$/Vs \\
\hline				   
$10^{19}$ & 01.18.35    & 167 $\pm$ 26 & 00.00.38 & 00.02.47 &  95 $\pm$ 40 \\
$10^{18}$ & 13.23.44    & 210 $\pm$ 46 & 00.07.39 & 00.02.17 & 198 $\pm$ 50 \\
$10^{17}$ & $\pm$ 145 h &              & 01.09.47 & 00.02.16 & 386 $\pm$ 75 \\
$10^{16}$ & $\pm$ 380 h &              & 03.04.02 & 00.03.40 & 400 $\pm$ 60 \\
$10^{15}$ & $\pm$ 80 h  &              & 00.38.43 & 00.06.39 & 412 $\pm$ 52 \\
$10^{14}$ & 09.56.23    & 452 $\pm$ 32 & 00.05.07 & 00.07.41 & 430 $\pm$ 43 \\
$10^{13}$ & 01.06.30    & 418 $\pm$ 38 & 00.00.32 & 00.07.55 & 420 $\pm$ 37 \\
$0$       & 00.06.37    & 408 $\pm$ 50 & 00.00.03 & 00.06.37 & 408 $\pm$ 50 \\
\hline
\end{tabular}
\end{center}
\label{Treleff}
\end{table}

In Table \ref{Treleff} the first column gives the concentration of
ionized impurities, the next three columns are data from the
isotropic code, and the last three columns are data from the
anisotropic code.  The errors indicate the 95 \% certainty level and
correspond to about twice the standard deviation.  In the case of an
isotropic choice of the angle after scattering, with 25 blocks of
1000 real scatterings execution time became too long for $n_{ii} =
10^{15}$ to $10^{17}$ cm$^{-3}$.  Therefore it was estimated from
runs with 2 blocks of 100 scatterings. We can see that in all cases
the anisotropic method was significantly faster than the isotropic --
the longest anisotropic execution time is 7 minutes 55 seconds,
whereas all but the $n_{ii}=0$ simulation using the isotropic code
took over 1 hour.

\begin{table}[h]
\caption{
A comparison of execution times and mobilities for simulations with 
the new anisotropic algorithm, with comparison to the tabulated 
mobilities from Sze[8].  These are for unstrained silicon at a 
temperature of 300 K and are averaged over $50 \times 5000$ 
scatterings.
}
\begin{center}
\begin{tabular}{|l|c|rcl|c|}
\hline
Conc.     & Exec. time & \multicolumn{3}{c|}{Mobility (simul.) }  & Mobility (Sze) \\
cm$^{-3}$ & hh.mm.ss   & \multicolumn{3}{c|}{cm$^2$/Vs} & cm$^2$/Vs \\
\hline			 	      
$10^{19}$ & 00.27.50   &   132 & $\pm$ &   13 &           \\
$10^{18}$ & 00.22.18   &   196 & $\pm$ &   21 &  150      \\
$10^{17}$ & 00.22.34   &   334 & $\pm$ &   24 &  300      \\
$10^{16}$ & 00.36.29   &   400 & $\pm$ &   19 &  430      \\
$10^{15}$ & 01.06.43   &   418 & $\pm$ &   14 &  460      \\
$10^{14}$ & 01.17.14   &   422 & $\pm$ &   14 &  470      \\
$10^{13}$ & 01.19.12   &   430 & $\pm$ &   12 &           \\
$0$       & 01.02.28   &   428 & $\pm$ &   14 &           \\
\hline
\end{tabular}
\end{center}
\label{Tcfsze}
\end{table}
As a final check, table \ref{Tcfsze} shows simulation results
compared with experimental values for the mobility in silicon as
found in Sze\cite{Sze-PSD}.  For more precision each simulation consisted
of 50 blocks of 5000 real scatterings, and the electric field was
kept at 5 kV/cm.  Note that again the stated errors correspond to a
95 \% certainty level, and that we have good agreement between our
code with the optimised ionized impurity scattering algorithm and the
values given by Sze.



\section{Conclusions}\label{Sconclusions}

We have demonstrated a new way of significantly reducing execution
times in simulations of systems involving Brooks-Herring ionized
impurity scattering.  This was done by introducing a carefully
calculated anisotropic scattering angle for these ionized impurity
processes, thus avoiding the usual problem of a large fraction of
inefficient overestimations.  Execution times of hole mobility
calculations in silicon showed that the speedup was strongly
dependent on impurity concentration, and generally well in excess of
a factor of 5, typically being of the order of 100.  The method is
applicable to any simulation in which Brookes-Herring ionized
impurity scattering is implemented, and it could be  usefully
generalised to other strongly anisotropic scattering processes, such
as polar optical phonon and piezoelectric scatterings.


\section*{Acknowledgements}
\label{acknowledgements}

\vspace{2mm}
\noindent

This work is funded by the European Commission via the 
program for Training and Mobility of Researchers.



\begin{thebibliography}{99}

\bibitem{Ridley-QPS} 
B.K. Ridley {\it Quantum Processes in Semiconductors}, 
 (Clarendon Press, Oxford, 1988).

\bibitem{Jacoboni-R-1983}
C. Jacoboni, L. Reggiani, Rev. Mod. Phys. B{\bf 55}, 645 (1983)

\bibitem{Konijn-RW-1993}
P. Konijn, T.G. van der Roer, F.P. Widdershoven,
Solid. State. Electron. {\bf 36}, 1579 (1993).

\bibitem{Kinsler-W-2001jap}
P. Kinsler, W.Th. Wenckebach,
J. Appl. Phys. {\bf 90}, 1692 (2001).

\bibitem{Dijkstra-W-1997jap}
J.E. Dijkstra, W.Th. Wenckebach, 
J. Appl. Phys. {\bf 81}, 1259 (1997).

\bibitem{Hinckley-S-1990}
J.M. Hinckley, J.Singh, 
Phys. Rev. B{\bf 41}, 2912 (1990).

\bibitem{Hinckley-S-1994}
J.M. Hinckley, J.Singh, 
J. Appl. Phys. {\bf 76}, 4192 (1994).

\bibitem{Sze-PSD} 
S.M. Sze, {\it Physics of Semiconductor Devices}, 
 (Wiley, New York, 1981).


\end{thebibliography}
\end{document}